\documentclass[preprint,prb,superscriptaddress]{revtex4}%
\usepackage{amsfonts}
\usepackage{amsmath}
\usepackage{amssymb}
\usepackage{graphicx}%
\newtheorem{theorem}{Theorem}
\newtheorem{acknowledgement}[theorem]{Acknowledgement}

\begin{document}
\title{Sonoluminescence and collapse dynamics of multielectron bubbles in helium}
\author{J. Tempere}
\affiliation{TFVS, Universiteit Antwerpen, Universiteitsplein 1, B2610 Antwerpen, Belgium}
\affiliation{Lyman Laboratory of Physics, Harvard University, Cambridge, Massachusetts}
\author{I. F. Silvera}
\affiliation{Lyman Laboratory of Physics, Harvard University, Cambridge, Massachusetts}
\author{S. Rekhi}
\affiliation{Lyman Laboratory of Physics, Harvard University, Cambridge, Massachusetts}
\thanks{Currently at Massachusetts Institute of Technology, Cambridge, Massachusetts}
\author{J. T. Devreese}
\affiliation{TFVS, Universiteit Antwerpen, Universiteitsplein 1, B2610 Antwerpen, Belgium}
\date{\today}

\begin{abstract}
Multielectron bubbles (MEBs) differ from gas-filled bubbles in that it is the
Coulomb repulsion of a nanometer thin layer of electrons that forces the
bubble open rather than the pressure of an enclosed gas. We analyze the
implosion of MEBs subjected to a pressure step, and find that despite the
difference in the underlying processes the collapse dynamics is similar to
that of gas-filled bubbles. When the MEB collapses, the electrons inside it
undergo strong accelerations, leading to the emission of radiation. This type
of sonoluminescence does not involve heating and ionisation of any gas inside
the bubble. We investigate the conditions necessary to obtain sonoluminescence
from multielectron bubbles and calculate the power spectrum of the emitted radiation.

\end{abstract}
\maketitle

\section{Introduction}

Multielectron bubbles (MEBs) are typically micron-sized cavities in liquid
helium, containing a nanometer thin layer of electrons on the inside of the
surface of the bubble. MEBs were first observed in experiments studying
electrons on the surface of liquid helium beneath which an anode was present
\cite{VolodinJETP26}. When the charge density of the 2D-layer of electrons
exceeds a critical value, a surface instability sets in \cite{GorkovJETP18}
and a bubble forms, carrying up to 10$^{8}$ electrons to the anode. The
electron system inside such a MEB is of particular interest, since it forms a
spherical two-dimensional electron gas whose density can be tuned over a wid e
range of values depending on the number of electrons in the bubble and the
pressure applied on the helium \cite{TemperePRL87,TempereEPJB32}.

Unlike in gas-filled bubbles, the force preventing the bubble from imploding
is not due to the pressure of a gas inside the bubble, but due to the Coulomb
repulsion between the electrons in the MEB. At zero external pressure, the
radius $R_{\text{b}}$ of the bubble is determined by the balance between the
Coulomb repulsion and the surface tension of the helium \cite{ShikinJETP27}.
When the pressure $p$ on the helium is nonzero, the bubble radius satisfies
the equation
\begin{equation}
\frac{e^{2}N^{2}}{4\pi\varepsilon R_{\text{b}}^{3}}=2pR_{\text{b}}+4\sigma,
\end{equation}
where $e$ is the electron charge, $N$ is the number of electrons in the MEB,
$\varepsilon$ is the dielectric constant of helium, and $\sigma$ is the
surface tension of helium. In the regime $pR_{\text{b}}/\sigma\gg1$ the bubble
radius scales as $p^{-1/4}$. In this paper we investigate what happens when a
bubble with radius $R_{\text{b}}(p=0)$ in equilibrium at zero pressure is
suddenly subjected to an increase in pressure. Its radius is larger than the
equilibrium radius at pressure $p$, $R_{\text{b}}(p=0)>R_{\text{b}}(p)$ and we
expect the bubble to shrink. In section II we derive and investigate the
equations of motion for the bubble dynamics, and in section III we apply these
equations to analyze the collapse dynamics of the multielectron bubble.

The fact that inside MEB's there are electrons (charged particles) instead of
neutral gas atoms necessarily has another interesting consequence: the
electrons in the compressing bubble are subject to accelerations, and
therefore they generate electromagnetic radiation. The breathing mode of the
bubble gives an estimate of the response time of the helium surface - which
typically is microseconds for a 10000 electron MEB. In that time span, the
bubble radius shrinks a distance of typically a micrometer. The velocity
corresponding to this is of the order of micrometer/microsecond or m/s. But
the acceleration this generates is on the order of a
micrometer/microsecond$^{2}$ which is quite large (10$^{6}$ m/s$^{2}$). As the
number of electrons increases, the initial bubble radius increases and the
maximum acceleration increases as well. We will show that ultrashort light
pulses can be produced by a sudden pressure increase in the helium surrounding
the MEB.

Explanations of sonoluminescence in gas-filled bubbles
\cite{GaitanJASA91,HillerPRL69,BrennerRMP74} have relied strongly on the
emissivity of the heated gas in the collapsing gas-filled bubble
\cite{HilgenfeldtNAT398}, as opposed to the mechanism of sonoluminescence
proposed here for multielectron bubbles. Nevertheless gas-filled bubbles are
known to get to temperatures where at least weak ionisation of the gas must
occur \cite{BrennerRMP74}, and electrons can emit light, being accelerated
through scattering from the neutral atoms. Although emission from accelerated
electrons can contribute to the sonoluminescence of gas-filled bubbles
\cite{HammerJMO48}, it's precise role is still disputed
\cite{HilgenfeldtNAT398}. For MEBs we propose that accelerating electrons form
the principal mechanism of sonoluminescence. Moreover the light emission is
expected to be coherent and coming from the bubble surface, whereas in
gas-filled bubbles the emission is incoherent black-body emission from the
interior of the bubble.

\section{Equations of motion for the bubble surface}

The bubble surface at time $t$ can be described by the function $R(\theta
,\phi,t)$ that gives the distance from the center of the bubble to the
surface, in the direction specified by the spherical coordinates $\theta,\phi
$. This function can be developed in spherical harmonics%
\begin{equation}
R(\theta,\phi,t)=R_{\text{b}}(t)+\sum_{\ell=1}^{\infty}\sum_{m=-\ell}^{\ell
}a_{\ell,m}(t)Y_{\ell,m}(\theta,\phi),
\end{equation}
where $R_{\text{b}}$ is the angle-averaged radius of the bubble. The
coefficients $a_{\ell m}$ describe the deformation of the bubble from its
spherical shape. To derive the equations of motion for the bubble surface, we
start from the bubble Lagrangian derived in \cite{TemperePRL87}:%
\begin{align}
\mathcal{L}  &  =2\pi\rho R_{\text{b}}^{3}\dot{R}_{\text{b}}^{2}-\left(
4\pi\sigma R_{\text{b}}^{2}+\frac{4\pi}{3}pR_{\text{b}}^{3}+\frac{e^{2}N^{2}%
}{2\varepsilon R_{\text{b}}}\right)  +\frac{\rho R_{\text{b}}^{3}}{2}%
\sum_{\ell,m}\frac{|\dot{a}_{\ell,m}|^{2}}{\ell+1}\nonumber\\
&  -\sum_{\ell,m}\left[  \frac{\sigma}{2}(\ell^{2}+\ell+2)+pR_{\text{b}}%
-\frac{e^{2}N^{2}}{8\pi\varepsilon R_{\text{b}}^{3}}\ell\right]  |a_{\ell
,m}|^{2}, \label{Lagr}%
\end{align}
where $\rho$ represents the mass density of liquid helium. For small amplitude
deformations of a spherical bubble, we know that $a_{\ell,m}(t)=a_{\ell
,m}(0)\times\exp\{i\omega_{\ell}t\}$ where the $\omega_{\ell}$ are the
spherical ripplon frequencies \cite{TemperePRL87}. These frequencies can be
obtained from (\ref{Lagr}) by putting $R_{\text{b}}(t)=R_{\text{eq}}$ where
$R_{\text{eq}}$ is the equilibrium radius:%
\begin{equation}
\omega_{\ell}=\sqrt{\frac{\ell+1}{\rho R_{\text{eq}}^{3}}\left[  \sigma
(\ell^{2}+\ell+2)+2pR_{\text{eq}}-\frac{e^{2}N^{2}}{4\pi\varepsilon
R_{\text{eq}}^{3}}\ell\right]  }. \label{ripfreq}%
\end{equation}
In the Lagrangian (\ref{Lagr}), we have neglected the influence of the
redistribution of the electrons along a deformed bubble surface. This effect
can be described by introducing the (spherical) plasma oscillation modes of
the electron system, and calculating the coupling between plasmons and
ripplons \cite{KliminSSC126}. In a recent study of this coupling Klimin
\textit{et al.} \cite{KliminSSC126} found that although a bubble deformation
does lead to a redistribution of charge along the surface, the resulting shift
of the ripplon frequencies is small. The reason for the weakness of the
ripplon-plasmon coupling is the difference in frequency of these modes: the
ripplon frequencies typically lie in the MHz-GHz range, whereas the plasmon
frequencies lie in the THz range.

The equation of motion for the radial component is
\begin{equation}
\frac{d}{dt}\frac{\partial\mathcal{L}}{\partial\dot{R}_{\text{b}}}%
=\frac{\partial\mathcal{L}}{\partial R_{\text{b}}},
\end{equation}
which yields
\begin{align}
\frac{3}{2}\left(  \frac{\dot{R}_{\text{b}}}{R_{\text{b}}}\right)  ^{2}%
+\frac{\ddot{R}_{\text{b}}}{R_{\text{b}}}  &  =-\frac{1}{\rho R_{\text{b}}%
^{4}}\left(  2\sigma R_{\text{b}}+pR_{\text{b}}^{2}-\frac{e^{2}N^{2}}%
{8\pi\varepsilon R_{\text{b}}^{2}}\right) \label{Req1}\\
&  +\frac{3}{8\pi R_{\text{b}}^{2}}\sum_{\ell,m}\frac{|\dot{a}_{\ell,m}|^{2}%
}{\ell+1}-\frac{1}{4\pi\rho R_{\text{b}}^{4}}\sum_{\ell,m}\left[
p-\frac{3e^{2}N^{2}}{8\pi\varepsilon R_{\text{b}}^{4}}\ell\right]  |a_{\ell
,m}|^{2}.\nonumber
\end{align}
The first line corresponds to the Rayleigh equation (with an additional
electronic term) describing the collapse of spherical bubbles
\cite{RayleighPM34}. The second line in (\ref{Req1}) contains additional terms
due to the deformation of the bubble.

The equation for the deformation components $\{a_{\ell,m}(t)\}$ is obtained
from
\begin{equation}
\frac{d}{dt}\frac{\partial\mathcal{L}}{\partial\dot{a}_{\ell,m}^{\ast}}%
=\frac{\partial\mathcal{L}}{\partial a_{\ell,m}^{\ast}},
\end{equation}
which leads to
\begin{equation}
\ddot{a}_{\ell,m}+3\frac{\dot{R}_{\text{b}}}{R_{\text{b}}}\dot{a}_{\ell
,m}=-\frac{\ell+1}{\rho R_{\text{b}}^{3}}\left[  \sigma(\ell^{2}%
+\ell+2)+2pR_{\text{b}}-\frac{e^{2}N^{2}}{4\pi\varepsilon R_{\text{b}}^{3}%
}\ell\right]  a_{\ell,m}. \label{aeq}%
\end{equation}
This equation is similar to the Plesset-Prosperetti equation
\cite{PlessetARFM9} in the theory of gas-filled bubbles in fluids. The set of
equations (\ref{Req1}) and (\ref{aeq}) need to be solved given appropriate
initial conditions.

Salomaa and Williams \cite{SalomaaPRL47} studied small-amplitude oscillations
of multielectron bubbles using a similar set of equations. They neglected the
contribution of the deformation amplitudes $\{a_{\ell,m}\}$'s in their
equation for the radial component $R_{\text{b}}(t)$. This approximation is
valid if $a_{\ell,m}\ll R_{\text{b}}$. They found that when the (time
averaged) bubble radius satisfies $R_{\text{eq}}<[N^{2}e^{2}/(16\pi
\sigma\varepsilon)]^{1/3}$, the $\ell=2$ deformation can grow indefinitely,
suggesting an instability of the bubble. In subsequent work
\cite{TemperePRL87} it was found that the $\ell=2$ ripplon frequency can
indeed vanish, corroborating the results of Salomaa and Williams. However, an
extension of ref. \cite{SalomaaPRL47} for large amplitude oscillation (i.e.
beyond the regime $a_{\ell,m}\ll R_{\text{b}}$) showed that there exists a
metastability barrier preventing the bubbles from fissioning
\cite{TemperePRB67}.

\section{Collapse of a multielectron bubble}

Having obtained the equations of motion for the bubble shape, we can proceed
to solve them for the particular case of an MEB subjected to a sudden increase
in pressure. We start with a bubble whose angle-averaged radius $R_{\text{b}}$
is equal to the equilibrium radius $R_{\text{eq}}(p_{0})$ at a given pressure
$p_{0}$ of the liquid helium surrounding the MEB. This bubble can be either
spherical ($\forall\ell,m:a_{\ell,m}(t=0)=0$) or deformed ($\exists
\ell,m:a_{\ell,m}(t=0)\neq0$). At time $t=0$ the pressure of the liquid helium
jumps from $p_{0}$ to a higher value $p$. The angle-averaged radius is then
larger than the equilibrium radius at pressure $p$ and the bubble will collapse.

\subsection{Spherical bubble}

First we consider the collapse of an undeformed bubble. Then all $a_{\ell m}%
$'s are zero and the dynamics are governed by the Rayleigh equation
\cite{RayleighPM34}
\begin{equation}
\frac{3}{2}\left(  \frac{\dot{R}_{\text{b}}}{R_{\text{b}}}\right)  ^{2}%
+\frac{\ddot{R}_{\text{b}}}{R_{\text{b}}}=-\frac{1}{\rho R_{\text{b}}^{4}%
}\left(  2\sigma R_{\text{b}}+pR_{\text{b}}^{2}-\frac{e^{2}N^{2}}%
{8\pi\varepsilon R_{\text{b}}^{2}}\right)  , \label{Req1b}%
\end{equation}
in which an additional term due to the Coulomb repulsion of the electrons
appears. This equation can be integrated once by multiplying left and right
hand side by $2R_{\text{b}}^{4}\dot{R}_{\text{b}}$, resulting in
\begin{equation}
\frac{d}{dt}\left(  R_{\text{b}}^{3}\dot{R}_{\text{b}}^{2}\right)  =-\frac
{1}{2\pi\rho}\frac{d}{dt}\left[  \left(  \frac{e^{2}N^{2}}{2\varepsilon
R_{\text{b}}}+4\pi\sigma R_{\text{b}}^{2}+\frac{4\pi}{3}pR_{\text{b}}%
^{3}\right)  \right]  ,
\end{equation}
so that
\begin{align}
\dot{R}_{\text{b}}^{2}  &  =-\frac{1}{2\pi\rho R_{\text{b}}^{3}}\left[
U(R_{\text{b}})-U(R_{\text{b}}(0))\right] \label{aa}\\
&  \text{where }U(R)=\frac{e^{2}N^{2}}{2\varepsilon R}+4\pi\sigma R^{2}%
+\frac{4\pi}{3}pR^{3}. \label{U(R)}%
\end{align}
This can again be integrated to yield
\begin{equation}
t(R)=%
%TCIMACRO{\dint \limits_{R_{\text{b}}(0)}^{R}}%
%BeginExpansion
{\displaystyle\int\limits_{R_{\text{b}}(0)}^{R}}
%EndExpansion
\left\{  \frac{1}{2\pi\rho r^{3}}\left[  U(R_{\text{b}}(0))-U(r)\right]
\right\}  ^{-1/2}dr. \label{Req2}%
\end{equation}
Note that the solution $R(t)$ is periodic, with the period given by
$T=t(R_{\text{turn}})$ where $R_{\text{turn}}$ is the classical turning point
in the potential $U(r)$. In figure \ref{fig1}, the time evolution of the
radius of a spherical $N=10000$ electron bubble subjected to a sudden increase
in pressure is shown. The top panel shows the pressure increase, and the
subsequent panels show the radius, the radial velocity and the radial
acceleration as a function of time divided by the period $T$. The initial
conditions are $\dot{R}_{\text{b}}(0)=0$ and $R_{\text{b}}(0)=1.06441$ $\mu$m
- this corresponds to the $N=10000$ electron bubble in equilibrium at zero
pressure. The periods are $T=0.795$ $\mu$s, $0.570$ $\mu$s and $0.462$ $\mu$s
for pressure steps of 1,2, and 3 kPa, respectively.

The radial acceleration shows a pronounced peak at the time when the MEB
radius is minimal (the `time of collapse'). Upon increasing the pressure, the
maximum radial acceleration grows and the oscillation of the radius becomes
more strongly anharmonic. There is no dissipation term in our equation for the
radius, so that the radius bounces back to its original value during the
periodic oscillation. Dissipation will reduce this value and will let the
bubble perform a damped oscillation around its new equilibrium radius
$R_{\text{eq}}(p)$. As can be seen from figure 1, the time evolution of the
radius of the MEB is very similar to that of a gas-filled bubble subjected to
a shock wave \cite{GaitanJASA91}. In the latter case, the restoring force
opposing the collapse is the increase in pressure of the gas trapped in the
bubble. In the present case, the bubble contains no gas (except helium atoms
at vapor pressure which is negligible at low temperatures), and it is the
increased Coulomb repulsion that gives rise to the restoring force.

\subsection{Deformed bubble}

For small-amplitude deformations, $a_{\ell,m}\ll R_{\text{b}}$, we can neglect
the $\left|  a_{\ell,m}\right|  ^{2}$ terms in the equation (\ref{Req1}) for
the bubble radius, and use the solution for $R_{\text{b}}(t)$ discussed in the
previous section directly in the equation (\ref{aeq}) for the deformation
amplitudes. This equation can then be written as%
\begin{equation}
\ddot{a}_{\ell,m}(t)+\lambda(t)\dot{a}_{\ell,m}(t)+\omega^{2}(t)a_{\ell
,m}(t)=0, \label{aeq2}%
\end{equation}
with the `drag coefficient':%
\begin{equation}
\lambda(t)=3\dot{R}_{\text{b}}(t)/R_{\text{b}}(t), \label{c1}%
\end{equation}
and the time dependent ripplon frequency%
\begin{equation}
\omega^{2}(t)=\frac{\ell+1}{\rho R_{\text{b}}^{3}(t)}\left[  \sigma(\ell
^{2}+\ell+2)+2pR_{\text{b}}(t)-\frac{e^{2}N^{2}}{4\pi\varepsilon R_{\text{b}%
}^{3}(t)}\ell\right]  . \label{c2}%
\end{equation}
The time dependence of these coefficients (\ref{c1}),(\ref{c2}) appears
through the time-dependence in $R_{\text{b}}$. The differential equation
(\ref{aeq2}) can be solved numerically. The main question of this subsection
concerns whether the deformations will vanish, grow without bound, or stay
finite during the collapse of the bubble. From the numerical analysis we find
that there are stable and unstable modes of deformation.

Note that even at a fixed pressure there are stable and unstable modes, as
mentioned at the end of section II. Moreover, at higher pressure, more modes
are unstable. When a pressure step $\Delta p$ is applied, a mode that was
stable at the initial pressure $p_{0}$ may be unstable at the final pressure
$p_{0}+\Delta p$.

An example of the time evolution of the amplitude of a stable mode is shown in
figure \ref{fig2}. For this figure, we start with an $N=10^{4}$ electron
bubble at $p_{0}=0$ that has an $\ell=25$ ripplon present. In this example,
$\ell$ is chosen arbitrarily, but sufficiently high so that $\omega_{\ell
}(p_{0}+\Delta p)>0$. Under the equilibrium conditions, the deformation
amplitude oscillates with the corresponding ripplon frequency. When the
pressure is suddenly increased by $\Delta p=10^{3}$ Pa, we find that the
oscillation perseveres, but that the oscillation amplitude is modulated with
the same periodicity as the bubble radius. When the bubble radius is minimal,
the oscillation amplitude of the deformation mode is maximal. This can be
understood by substituting the ansatz $a_{\ell}(t)=A(t)e^{i\omega t}$ in the
equation of motion (\ref{aeq2}). Collecting the imaginary parts of this
equation, we arrive at
\begin{equation}
2\frac{\dot{A}(t)}{A(t)}+3\frac{\dot{R}_{\text{b}}}{R_{\text{b}}}=0,
\end{equation}
from which
\begin{equation}
A(t)=CR_{\text{b}}^{-3/2}(t).
\end{equation}
The constant of integration is given by $C=A(0)R_{\text{b}}^{3/2}(0)$. This
solution is shown in figure \ref{fig2} by the dashed curve. It fits well with
the envelope of the full numerical solution (full curve), even though we
started with a rather large deformation amplitude (10\% of the radius) in
order to get a clear figure. For stable modes, we thus find that during the
collapse of a deformed bubble the deformations do not vanish. The deformation
amplitude grows, but is bounded at all times.

An example of the time evolution of an unstable mode is shown in figure
\ref{fig3}. As in figure \ref{fig2} the initial state is that of an MEB with
$N=10^{4}$ electrons at pressure zero, subjected to a pressure increase of
$10^{3}$ Pa. The different curves in figure \ref{fig3} represent different
initial conditions, as listed in the legend. We chose the $\ell=2$ mode
because it is known to have a vanishing frequency at positive pressures and
thus may lead to instabilities. The numerical solution of (\ref{aeq2}) shows
that the deformation amplitude $a_{2}$ does indeed grow to a point where it
becomes comparable to the bubble radius. This indicates that the collapsing
MEB\ in helium becomes unstable and may break up, in contrast to gas-filled
bubbles in cryogenic liquids \cite{WilliamsPhysB284}. Nevertheless, at the
time of smallest radius, the deformation amplitude is still smaller than the
bubble radius. The point of instability ($a_{2}=R_{\text{b}}$) is reached
after the collapse. So, although the bubble may not survive long after the
first collapse, it will reach the regime where the radial acceleration is
maximal one time. How long the bubble survives after it has reached its
smallest radius, depends strongly on the initial conditions of the
deformation. To study the exact dynamics of the instability (the possible
fissioning), the full set of equations (\ref{Req1}),(\ref{aeq}) needs to be
solved. This lies beyond the scope of this paper as we want to investigate the
collapse dynamics and luminescence during collapse, but not the possible
breaking up of bubbles during the explosion that follows after the implosion.

\section{Light radiation from an imploding MEB}

\subsection{Incoherent emission from a spherical bubble}

\begin{figure}
[ptb]
\begin{center}
\includegraphics[
height=4.3106in,
width=3.3372in
]%
{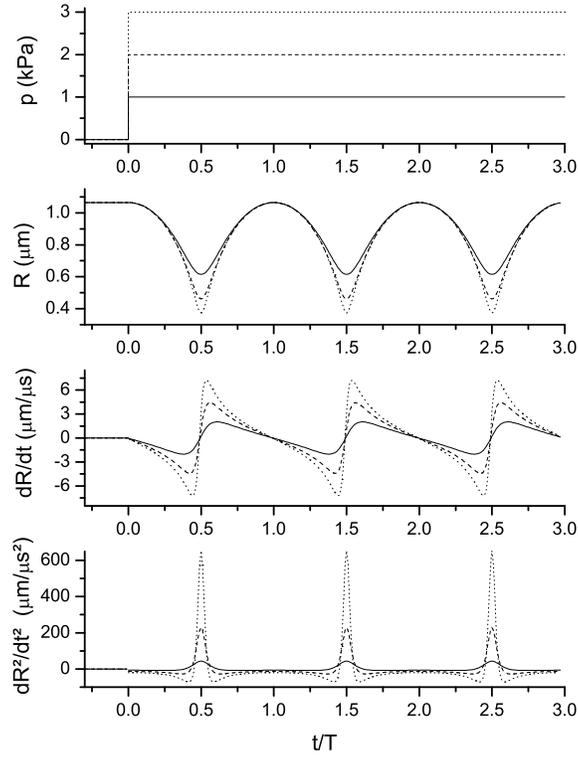}%
\caption{A multielectron bubble with $N=10^{4}$ electrons at zero external
pressure, is subjected at time $t=0$ to a pressure increase of $\Delta
p=1,2,3$ kPa (full, dashed and dotted curves, respectively). In this figure,
the panels from top to bottom show the pressure, the bubble radius, the radial
velocity and the radial acceleration as a function of time. All these
quantities are cyclic with period $T=0.795$ $\mu$s, $0.570$ $\mu$s and $0.462$
$\mu$s for pressure steps of 1,2, and 3 kPa, respectively.}%
\label{fig1}%
\end{center}
\end{figure}

\begin{figure}
[ptb]
\begin{center}
\includegraphics[
height=3.3572in,
width=4.241in
]%
{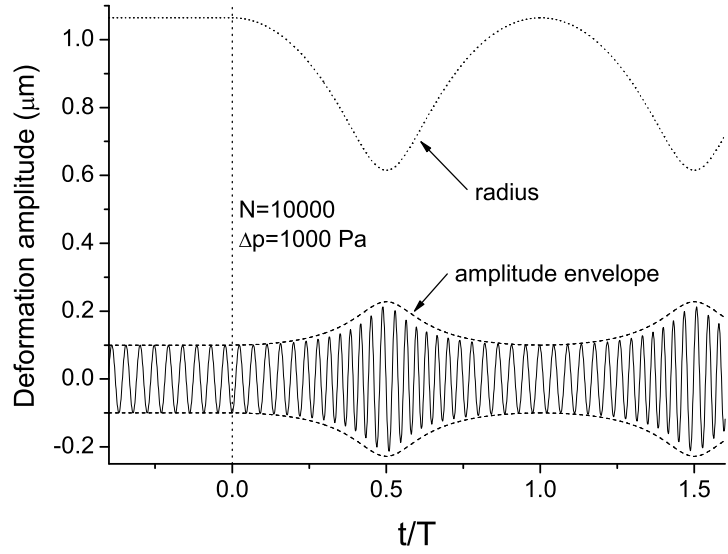}%
\caption{The time evolution of the amplitude of deformation associated with
the $\ell=25$ ripplon is shown (full curve) for an $N=10^{4}$ electron bubble
subjected to a pressure increase of $1$ kPa at time $t=0$. The dashed curve
shows the analytical estimate for the the envelope of the oscillating
deformation amplitude. The dotted curve depicts the radius of the bubble.}%
\label{fig2}%
\end{center}
\end{figure}

\begin{figure}
[ptb]
\begin{center}
\includegraphics[
height=3.288in,
width=4.2281in
]%
{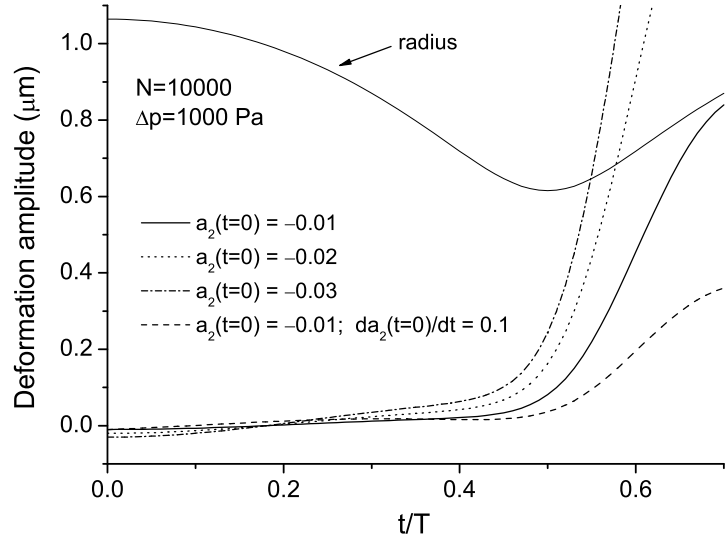}%
\caption{The time evolution of an unstable ($\ell=2$) mode of an $N=10^{4}$
bubble subjected to a pressure increase of $p=10^{3}$ kPa. The deformation
amplitude grows larger than the bubble radius (so that the bubble is unstable)
some time after collapse (i.e. after the bubble radius reaches its minimum),
depending on the initial conditions.}%
\label{fig3}%
\end{center}
\end{figure}

\begin{figure}
[ptb]
\begin{center}
\includegraphics[
height=3.6365in,
width=4.7444in
]%
{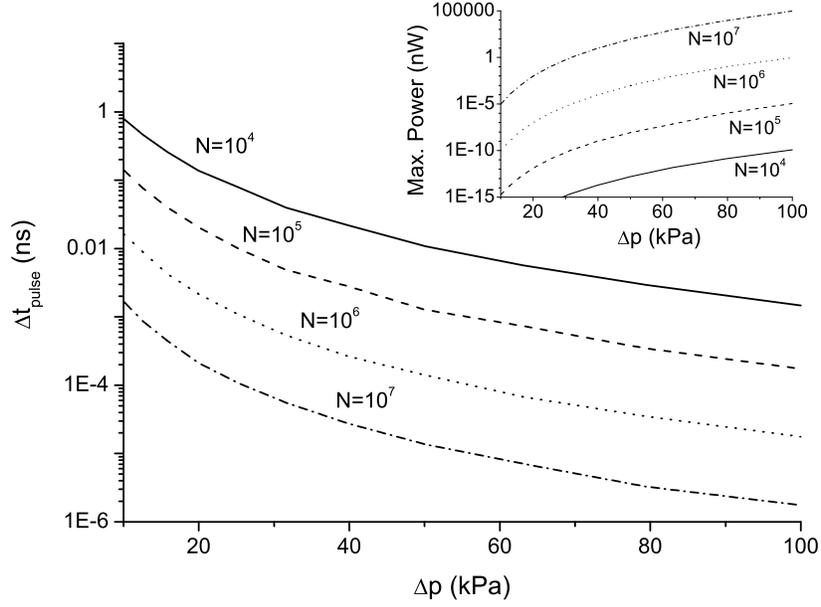}%
\caption{The width at half maximum $\Delta t_{\text{pulse}}$ of the peak in
the radial acceleration of the collapsing bubble as it reaches its smallest
radius is shown as a function of the pressure step causing the collapse, for
different values of the number of electrons in the MEB. In the inset, the
maximum power of the emitted radiation pulse is shown as a function of the
pressure step and the number of electrons.}%
\label{fig4}%
\end{center}
\end{figure}

\begin{figure}
[ptb]
\begin{center}
\includegraphics[
height=3.3578in,
width=4.0465in
]%
{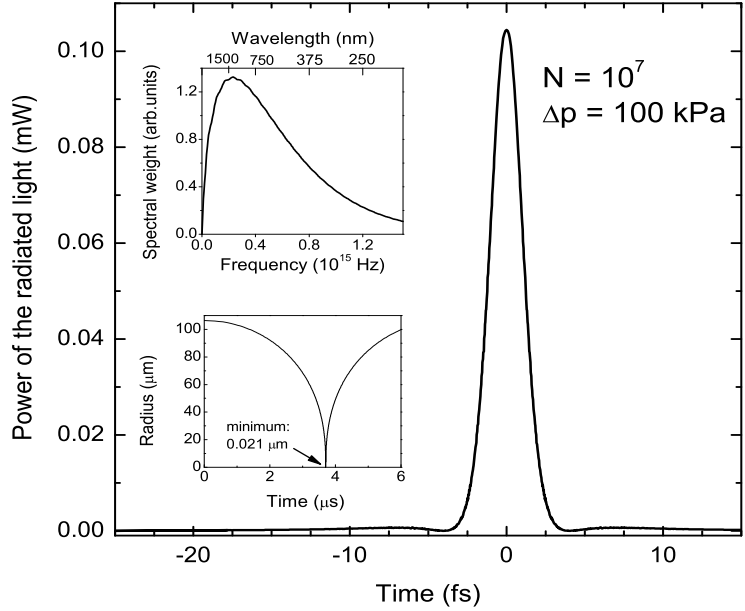}%
\caption{For very large MEBs (10$^{7}$ electrons) subjected to a large
pressure step (100 kPa), the emitted pulse of radiation calculated with (
\ref{Req1}),(\ref{rad}) can become extremely short and reach 100 $\mu$W peak
power. This also results in a broad spectrum of the emitted radiation, as
shown in the upper inset. The lower inset shows the bubble radius as a
function of time (cf fig. 1).}%
\label{fig5}%
\end{center}
\end{figure}

The electric radiation field by an accelerated charge is given by
\begin{equation}
\mathbf{E}(\mathbf{r},\mathbf{R},t)=\dfrac{q}{c^{2}}\left.  \dfrac
{\mathbf{n}\times\lbrack\mathbf{n}\times\mathbf{\ddot{R}}(t)]}{|\mathbf{r}%
-\mathbf{R}(t)|}\right|  _{\text{ret}}, \label{rad}%
\end{equation}
where $\mathbf{R}(t)$ gives the trajectory of the charge $q$, $\mathbf{r}$ is
the point in which we want to evaluate the electric radiation field,
$\mathbf{n}$ is the unit vector along $\mathbf{r}-\mathbf{R}(t)$, and $c$ is
the speed of light in vacuum. Note that there is a non-radiative part of the
electric field (of order $[\mathbf{r}-\mathbf{R}(t)]^{-2}$ and higher). This
is usually neglected at large distances from the accelerated electron.
However, if the radiative field is cancelled, for example by symmetry
considerations, then these higher order fields come into play so that the
Maxwell equations remain satisfied. The subscript ``ret'' means that the
electric field at time $t$ is generated by the accelerating electron at a time
$t^{\prime}$ earlier by the delay it takes the light to reach \textbf{r}. In
our situation, this retardation is of no consequence. From figure \ref{fig2},
it can be seen that though the radial velocity of the electrons is fairly
high, it is not relativistic (the acceleration, be it large, only takes place
in a very short time scale). The power radiated per unit of solid angle per
electron is%

\begin{equation}
\frac{dP}{d\Omega}=\frac{e^{2}}{4\pi c^{3}}|\mathbf{n}\times\lbrack
\mathbf{n}\times\mathbf{\ddot{R}}]|^{2}=\frac{e^{2}}{4\pi c^{3}}|\ddot{R}%
|^{2}\sin^{2}\theta,
\end{equation}
with $\theta$ the angle between the direction of the acceleration and
\textbf{n}. Thus the maximum radiated power is%
\begin{equation}
P=\frac{2Ne^{2}}{3c^{3}}|\ddot{R}|^{2}. \label{pow}%
\end{equation}
We first consider a light pulse that might be emitted by a single electron in
an MEB. The light pulse intensity is peaked at the instant when the bubble
reaches its smallest radius, stops imploding and starts expanding again. At
this point, the radial acceleration $\left\vert \ddot{R}\right\vert $ is
largest. This is similar to what happens in sonoluminescence in gas-filled
bubbles. For bubbles with 10000 electrons, and a pressure step of 3 kPa, a
pulse of roughly 50 ns is emitted, as can be seen from figure \ref{fig1},
bearing in mind that in that case $T=0.462$ $\mu$s. Increasing the pressure
step will shorten the length of the pulse, and increase the maximum
acceleration reached (see figure \ref{fig1}). In figure \ref{fig4} we show the
width $\Delta t_{\text{pulse}}$ at half maximum of the peak in the radial
acceleration, as a function of pressure for different numbers of electrons.
Large MEBs ($N=10^{7}$), such as those created in the experiments reported in
Refs. \cite{VolodinJETP26}, have a very high compressibility resulting in
ultrashort pulses and very large radial accelerations. Figure \ref{fig5} shows
the power as a function of time (and frequency, in the inset) for an extreme
case: an $N=10^{7}$ electron bubble subjected to a pressure step of 101.3 kPa.
The ultrashort pulse reaches a maximum power of 100 $\mu$W (i.e. $10^{-11}$ W
per electron). Of course, one can ask whether for such an extreme case, the
equation of motion (\ref{Req1b}) remains valid up to the time that the
smallest radius is reached. If the surrounding helium is heated up by the
radiation, or if the electrons penetrate into the helium, then the bubble
implosion may have different dynamics. Nevertheless, it seems safe to expect
that equation (\ref{Req1}) remains valid for a large portion of the collapse,
and that for not so extreme cases the qualitative results hold. These results
are that if the number of electrons or the pressure step is increased, the
pulse duration shortens, the emitted spectrum shifts to higher frequencies,
and the maximum power of the pulse increases.

\subsection{Sonoluminescence from deformed bubbles}

The results of the previous subsection were for one accelerating electron, or
for a collection of electrons emitting radiation incoherently. If the electric
field of all the electrons is added coherently, then for an undeformed bubble
the radiation of the collapsing shell of electrons will vanish due to the
spherical symmetry, as we shall prove.

We investigate the case where one of the deformation amplitudes $a_{\ell,m}$
is nonzero. In studying modes of vibration where $m\neq0$, the restriction
$a_{\ell,-m}=a_{\ell,m}^{\ast}$ applies because the surface $R(\Omega,t)$ has
to be described by a real function, and thus
\begin{align}
\mathbf{R}(\Omega,t)  &  =\left\{  R_{\text{b}}(t)+2\operatorname{Re}\left[
a_{\ell,m}(t)Y_{\ell,m}(\Omega)\right]  \right\}  \mathbf{n}_{\Omega}\\
&  =R_{\text{b}}(t)\left[  1+\tilde{a}_{\ell,m}(t)P_{\ell}^{m}(\cos\theta
)\cos(m\phi)\right]  \mathbf{n}_{\Omega}, \label{Rdef}%
\end{align}
where $P_{\ell}^{m}$ is the associated Legendre function, $\mathbf{n}_{\Omega
}$ is the unit vector normal to the surface and the rescaled deformation
amplitude $\tilde{a}_{\ell,m}$ is given by%
\begin{equation}
\tilde{a}_{\ell,m}(t)=2\dfrac{a_{\ell,m}(t)}{R_{b}(t)}\sqrt{\dfrac{2\ell
+1}{4\pi}\dfrac{(\ell-m)!}{(\ell+m)!}}.
\end{equation}
We evaluate the electric field at a point
\begin{equation}
\mathbf{r}=r\left(  \mathbf{e}_{x}\sin\alpha\sin\beta+\mathbf{e}_{y}\sin
\alpha\cos\beta+\mathbf{e}_{z}\cos\alpha\right)  ,
\end{equation}
where we assume that $r\gg R$, i.e. the radiation field is evaluated at a
large distance from the bubble. Furthermore, we will assume that $\ddot
{a}_{\ell,m}\ll\ddot{R}_{\text{b}}$, which is valid when the bubble is near
its minimum radius, and when the initial deformation satisfies $a_{\ell
,m}(t=0)\ll R_{\text{b}}(t=0)$. Substitution of (\ref{Rdef}) in (\ref{rad})
results in%
\begin{align}
\mathbf{E}(r  &  \gg R)\approx\dfrac{Ne}{c^{2}}\dfrac{\ddot{R}_{\text{b}}%
(t)}{r}\times\int d\Omega\left\{  \dfrac{R(\Omega,t)}{R_{\text{b}}(t)}%
\Theta(\Omega)\text{ }\mathbf{e}_{\mathbf{r}}-\dfrac{\mathbf{R}(\Omega
,t)}{R_{\text{b}}(t)}\right\} \\
&  =\dfrac{Ne}{c^{2}}\dfrac{\ddot{R}_{\text{b}}(t)}{rR_{\text{b}}(t)}%
\times\int d\Omega\text{ }\left[  1+\tilde{a}_{\ell,m}(t)P_{\ell}^{m}%
(\cos\theta)\cos(m\phi)\right]  \left[  \Theta(\theta,\phi)\mathbf{e}%
_{\mathbf{r}}-\mathbf{e}_{\mathbf{R}}\right]  ,
\end{align}
where the angular function is given by%
\begin{equation}
\Theta(\theta,\phi)=\cos\alpha\cos\theta+\sin\alpha\cos\beta\sin\theta\cos
\phi+\sin\alpha\sin\beta\sin\theta\sin\phi.
\end{equation}
The components of the electric field are, for $r\gg R$%
\begin{align}
\left(
\begin{array}
[c]{c}%
E_{x}\\
E_{y}\\
E_{z}%
\end{array}
\right)   &  \approx\dfrac{q}{c^{2}}\dfrac{\tilde{a}_{\ell,m}(t)\ddot
{R}_{\text{b}}(t)}{r}\times%
%TCIMACRO{\dint \limits_{0}^{\pi}}%
%BeginExpansion
{\displaystyle\int\limits_{0}^{\pi}}
%EndExpansion
d\theta\text{ sin}\theta\text{ }P_{\ell}^{m}(\cos\theta)%
%TCIMACRO{\dint \limits_{0}^{2\pi}}%
%BeginExpansion
{\displaystyle\int\limits_{0}^{2\pi}}
%EndExpansion
d\phi\text{ }\cos(m\phi)\nonumber\\
&  \times\left(
\begin{array}
[c]{c}%
\Theta(\theta,\phi)\sin\alpha\cos\beta-\sin\theta\cos\phi\\
\Theta(\theta,\phi)\sin\alpha\sin\beta-\sin\theta\sin\phi\\
\Theta(\theta,\phi)\cos\alpha-\cos\theta
\end{array}
\right)  .
\end{align}
From the integration over the $\phi$ coordinate, it is clear that only the
$m=0,\pm1$ terms result in a nonzero radiation field. The integration over the
$\theta$ angle finally yields%
\begin{align}
\left(
\begin{array}
[c]{c}%
E_{x}\\
E_{y}\\
E_{z}%
\end{array}
\right)   &  \approx\sqrt{\dfrac{4\pi}{3}}\dfrac{Ne}{c^{2}}\dfrac{\ddot
{R}_{\text{b}}(t)}{rR_{\text{b}}(t)}\nonumber\\
&  \times\left(
\begin{array}
[c]{c}%
(\cos\alpha\sin\alpha\cos\beta)a_{1,0}+(1-\sin^{2}\alpha\cos^{2}\beta
)a_{1,1}\\
\left(  \cos\alpha\sin\alpha\sin\beta\right)  a_{1,0}-(\sin^{2}\alpha\sin
\beta\cos\beta)a_{1,1}\\
(\sin^{2}\alpha)a_{1,0}-(\sin\alpha\cos\alpha\cos\beta)a_{1,1}%
\end{array}
\right)  .
\end{align}
Note that only the $\ell=1$ mode contributes to the radiation field. This
proves our earlier statement that the radiation fields of electrons on an
accelerating \textit{spherical} shell cancel out. The emitted power is then
given by%
\begin{equation}
P(\alpha,\beta)=N^{2}P_{0}\left[  \left(
\begin{array}
[c]{c}%
\sin\alpha\cos\beta\cos\alpha\\
\sin\alpha\sin\beta\cos\alpha\\
-\sin^{2}\alpha
\end{array}
\right)  \frac{a_{1,0}}{R_{\text{b}}}-\left(
\begin{array}
[c]{c}%
\sin^{2}\alpha\cos^{2}\beta-1\\
\sin^{2}\alpha\cos\beta\sin\beta\\
\cos\alpha\sin\alpha\cos\beta
\end{array}
\right)  \frac{a_{1,1}}{R_{\text{b}}}\right]  ^{2}, \label{pow2}%
\end{equation}
with $P_{0}=(4\pi/3)e^{2}\ddot{R}_{\text{b}}^{2}/c^{3}$. Using typical values
$N=10^{6},a_{\ell m}\approx0.01$ $R_{\text{b}}$, $p_{0}=0$ and $\Delta p=100$
kPa, we find a maximum power of the order of $100$ $\mu$W (for a pulse length
of the order of 10 fs). Although this may seem a lot, because of the short
pulse time only a small amount of energy is emitted, corresponding to 10-100
photons with frequency $1/(10$ fs). This will make the detection of the light
difficult in practice. Photons of successive collapse-expansion cycles may be
collected, but dissipation may quickly dampen the oscillations. If the bubble
oscillations after a single pressure step dampen out quickly, repeating the
pressure step using an ultrasonic wave of e.g. 1 KHz can increase the number
of photons to be collected. Another possibility may be to work with larger
bubbles since the total number of electrons affects very strongly the maximum
peak power, through the coherency factor $N^{2}$ in (\ref{pow2}) and the
strong dependence of $\ddot{R}_{\text{b}}$ on $N$. For example, a $N=10^{5}$
electron bubble subjected to the same $\Delta p=100$ kPa pressure step has a
maximum peak power of 10$^{-2}$ $\mu$W (for $a_{\ell m}\approx0.1$
$R_{\text{b}}$). In the extreme case of large bubbles $(N>10^{7}$) subjected
to large pressure steps ($\Delta p$ of the order of 10$^{5}$ Pa), the power
radiated can become large enough to affect the collapse dynamics
significantly, and expression (\ref{pow2}) is no longer valid. Two factors
will act to reduce the maximum power: incoherency effects (such as those
caused by the finite time needed for a pressure wave to travel across the
bubble) and the smallness of the deformations (typically, we assumed $a_{\ell
m}\ll R_{\text{b}}$).

\begin{figure}
[ptb]
\begin{center}
\includegraphics[
height=5.0462in,
width=5.9525in
]%
{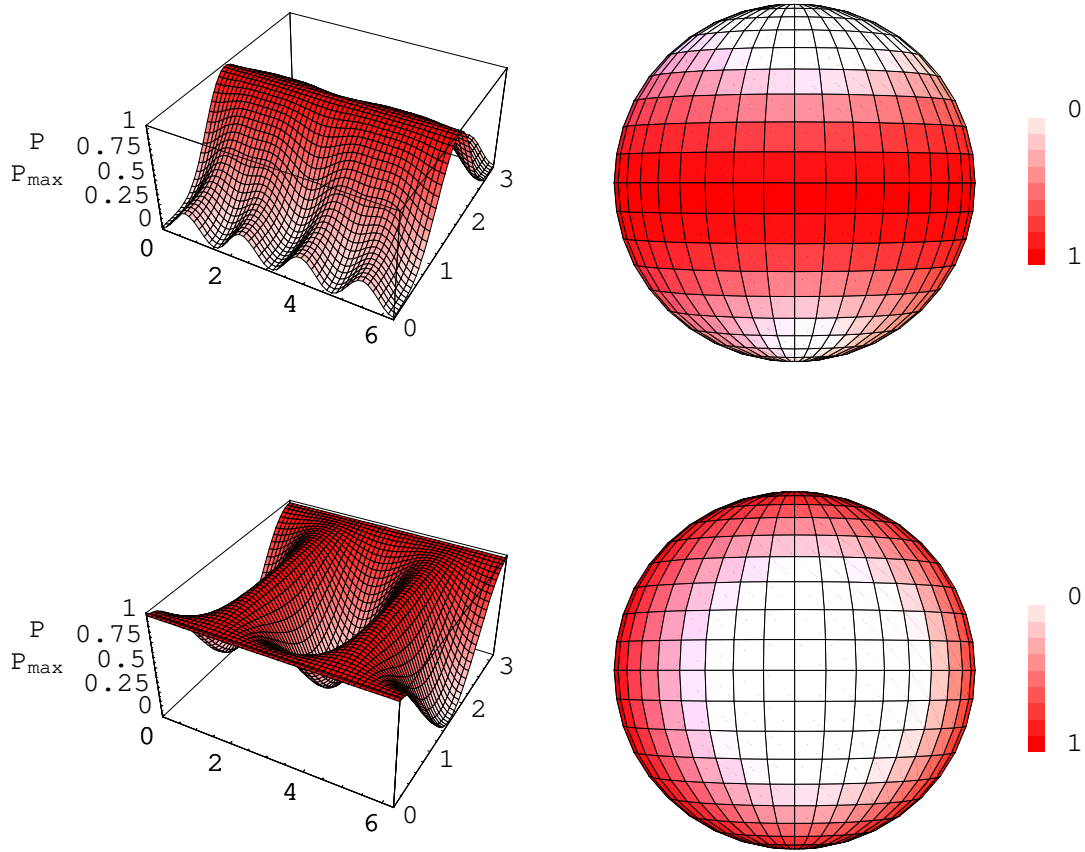}%
\caption{The maximum power (expression\ (\ref{pow2})) of the radiation from a
collapsing deformed bubble is shown as a function of the spherical angles
$\theta$ and $\phi$. The power is expressed relative to maximum power as a
function of the angles. The top two figures are for the $a_{1,0}\neq0$,
$a_{1,1}=0$ case and the bottom two figures for the $a_{1,0}=0$, $a_{1,1}%
\neq0$ case. The illustrations on the right show the regions on the sphere
(with the $z$-axis in the vertical direction) where radiation is emitted
(shaded regions) in both cases. }%
\label{fig6}%
\end{center}
\end{figure}

The last factor in (\ref{pow2}) is the geometrical factor giving the
angle-dependence of the radiation and relating its strength to the deformation
amplitude. The collapsing deformed bubble will radiate anisotropically. Figure
\ref{fig6} shows the angular dependence of the radiated power. The
$(\ell,m)=(1,0)$ mode will emit most strongly in the equatorial plane of the
bubble, whereas the $(\ell,m)=(1,\pm1)$ modes will emit most strongly along meridians.

\section{Discussion and conclusion}

When an MEB is subjected to a sudden increase in pressure $p_{0}\rightarrow
p_{0}+\Delta p$, its radius is larger than the new equilibrium radius
$R_{\text{eq}}(p_{0})>R_{\text{eq}}(p_{0}+\Delta p)$ and it shrinks to a final
radius that is smaller than the new equilibrium radius $R_{\text{eq}}%
(p_{0}+\Delta p)$, after which it expands again. This oscillatory motion of
collapses and expansions occurs on a time scale of microseconds, but the
oscillation becomes greatly anharmonic, in that most of the reduction in
radius occurs in a small fraction of the total cycle. This leads to large
accelerations when the radius of the bubble is smallest. When deformations are
present, the normal modes of deformation with a ripplon frequency larger than
the inverse time scale of the collapse are stable. Modes with small ripplon
frequencies (such as the $\ell=2$ quadrupole oscillation modes) grow unbounded
and can lead to instabilities and fissioning of the collapsing bubble.
However, we find that at the time when the bubble has reached its smallest
radius, and the radial acceleration is maximal, the deformation mode is not
yet diverging. This conclusion is important for the possibility to observe
sonoluminescence of the collapsing MEBs. In a purely spherical bubble,
coherent emission of radiation by the accelerated electrons cancel out due to
symmetry requirements. Thus, to produce a radiation field, the collapsing
bubble should be deformed. We find that in a collapsing bubble where the
$\ell=1$ mode of deformation is present, the radiation fields of the electrons
do not cancel. One might wonder how this particular mode can be excited so
that sonoluminescence can be generated in an experiment. It turns out that it
is not necessary to artificially excite this mode: Plesset \cite{PlessetARFM9}
showed that gas-filled bubbles that implode near a surface naturally develop a
deformation corresponding exactly to the $\ell=1$ mode, and we expect that
this surface proximity effect will also be present for imploding multielectron bubbles.

In the derivation, it was assumed that the temperature is low enough so that
the vapor pressure of helium inside the MEB is negligible. It is worth noting
that if helium vapor is present inside the bubble, the pressure of this vapor
will increase as the bubble collapses and a mist of helium may be formed that
can affect the dynamics of the bubble collapse \cite{TakemuraJSME37}.

In summary, we have investigated the dynamics of collapsing multielectron
bubbles in liquid helium and the possibility that imploding MEBs emit
radiation. We find that these bubbles behave in a similar manner as gas-filled
bubbles, even though the fundamental mechanism involved in the bubble dynamics
is different. In MEBs the force counteracting the collapse bubble is Coulomb
repulsion, whereas in gas-filled bubbles this force is due to the gas pressure
inside the bubble. The collapse of a MEB subjected to a pressure step leads to
the emission of radiation due to the acceleration of the shell of electrons at
the bubble surface, whereas sonoluminescence in gas-filled bubbles is due to a
more complex combined effect of heated gas emissivity and bremsstrahlung.

\begin{acknowledgement}
Discussions with D. Lohse are gratefully acknowledged. J. T. is supported
financially by the FWO-Flanders. This research has been supported by the
Department of Energy, Grant DE-FG02-ER45978, and by the GOA BOF UA 2000,
NOI\ BOF UA 2004, IUAP, the FWO-V projects Nos. G.0435.03, G.0274.01, G.0306.00.
\end{acknowledgement}

\bigskip%

%TCIMACRO{\FRAME{ftbpFU}{3.3372in}{4.3106in}{0pt}{\Qcb{A multielectron bubble
%with $N=10^{4}$ electrons at zero external pressure, is subjected at time
%$t=0$ to a pressure increase of $\Delta p=1,2,3$ kPa (full, dashed and dotted
%curves, respectively). In this figure, the panels from top to bottom show the
%pressure, the bubble radius, the radial velocity and the radial acceleration
%as a function of time. All these quantities are cyclic with period $T=0.795$
%$\mu$s, $0.570$ $\mu$s and $0.462$ $\mu$s for pressure steps of 1,2, and 3
%kPa, respectively.}}{\Qlb{fig1}}{figure1.eps}%
%maintain-aspect-ratio TRUE;  display "ICON";  valid_file "F";
%width 3.3372in;  height 4.3106in;  depth 0pt;  original-width 3.2915in;
%original-height 4.2583in;  cropleft "0";  croptop "1";  cropright "1";
%cropbottom "0";  filename 'figure1.eps';file-properties "XNPEU";}} }%
%BeginExpansion
%EndExpansion
%

%TCIMACRO{\FRAME{ftbpFU}{4.241in}{3.3572in}{0pt}{\Qcb{The time evolution of the
%amplitude of deformation associated with the $\ell=25$ ripplon is shown (full
%curve) for an $N=10^{4}$ electron bubble subjected to a pressure increase of
%$1$ kPa at time $t=0$. The dashed curve shows the analytical estimate for the
%the envelope of the oscillating deformation amplitude. The dotted curve
%depicts the radius of the bubble.}}{\Qlb{fig2}}{figure2.eps}%
%maintain-aspect-ratio TRUE;  display "ICON";  valid_file "F";  width 4.241in;
%height 3.3572in;  depth 0pt;  original-width 4.19in;
%original-height 3.3105in;  cropleft "0";  croptop "1";  cropright "1";
%cropbottom "0";  filename 'figure2.eps';file-properties "XNPEU";}}}%
%BeginExpansion
%EndExpansion
%

%TCIMACRO{\FRAME{ftbpFU}{4.2281in}{3.288in}{0pt}{\Qcb{The time evolution of an
%unstable ($\ell=2$) mode of an $N=10^{4}$ bubble subjected to a pressure
%increase of $p=10^{3}$ kPa. The deformation amplitude grows larger than the
%bubble radius (so that the bubble is unstable) some time after collapse (i.e.
%after the bubble radius reaches its minimum), depending on the initial
%conditions.}}{\Qlb{fig3}}{figure3.eps}{\special{ language "Scientific Word";
%type "GRAPHIC";  maintain-aspect-ratio TRUE;  display "ICON";
%valid_file "F";  width 4.2281in;  height 3.288in;  depth 0pt;
%original-width 4.1779in;  original-height 3.243in;  cropleft "0";
%filename 'figure3.eps';file-properties "XNPEU";}}}%
%BeginExpansion
%EndExpansion
%

%TCIMACRO{\FRAME{ftbpFU}{4.7444in}{3.6365in}{0pt}{\Qcb{The width at half
%maximum $\Delta t_{\text{pulse}}$ of the peak in the radial acceleration of
%the collapsing bubble as it reaches its smallest radius is shown as a function
%of the pressure step causing the collapse, for different values of the number
%of electrons in the MEB. In the inset, the maximum power of the emitted
%radiation pulse is shown as a function of the pressure step and the number of
%electrons.}}{\Qlb{fig4}}{figure4.eps}{\special{ language "Scientific Word";
%type "GRAPHIC";  maintain-aspect-ratio TRUE;  display "ICON";
%valid_file "F";  width 4.7444in;  height 3.6365in;  depth 0pt;
%original-width 4.6916in;  original-height 3.589in;  cropleft "0";
%filename 'figure4.eps';file-properties "XNPEU";}}}%
%BeginExpansion
%EndExpansion
%

%TCIMACRO{\FRAME{ftbpFU}{4.0465in}{3.3578in}{0pt}{\Qcb{For very large MEBs
%(10$^{7}$ electrons) subjected to a large pressure step (100 kPa), the emitted
%pulse of radiation calculated with ( \ref{Req1}),(\ref{rad}) can become
%extremely short and reach 100 $\mu$W peak power. This also results in a broad
%spectrum of the emitted radiation, as shown in the upper inset. The lower
%inset shows the bubble radius as a function of time (cf fig. 1).}}{\Qlb{fig5}%
%}{figure5.eps}{\special{ language "Scientific Word";  type "GRAPHIC";
%maintain-aspect-ratio TRUE;  display "ICON";  valid_file "F";
%width 4.0465in;  height 3.3578in;  depth 0pt;  original-width 3.9972in;
%original-height 3.3105in;  cropleft "0";  croptop "1";  cropright "1";
%cropbottom "0";  filename 'figure5.eps';file-properties "XNPEU";}}}%
%BeginExpansion
%EndExpansion
%

%TCIMACRO{\FRAME{ftbpFU}{5.9525in}{5.0462in}{0pt}{\Qcb{The maximum power
%(expression\ (\ref{pow2})) of the radiation from a collapsing deformed bubble
%is shown as a function of the spherical angles $\theta$ and $\phi$. The power
%is expressed relative to maximum power as a function of the angles. The top
%two figures are for the $a_{1,0}\neq0$, $a_{1,1}=0$ case and the bottom two
%figures for the $a_{1,0}=0$, $a_{1,1}\neq0$ case. The illustrations on the
%right show the regions on the sphere (with the $z$-axis in the vertical
%direction) where radiation is emitted (shaded regions) in both cases. }%
%}{\Qlb{fig6}}{figuur6.gif}{\special{ language "Scientific Word";
%type "GRAPHIC";  maintain-aspect-ratio TRUE;  display "ICON";
%valid_file "F";  width 5.9525in;  height 5.0462in;  depth 0pt;
%original-width 12.7292in;  original-height 10.7816in;  cropleft "0";
%filename 'figure6.eps';file-properties "XNPEU";}}}%
%BeginExpansion
%EndExpansion

\end{document}